\documentclass[sigconf]{acmart}
\usepackage{subcaption}

\AtBeginDocument{%
  \providecommand\BibTeX{{%
    \normalfont B\kern-0.5em{\scshape i\kern-0.25em b}\kern-0.8em\TeX}}}

\copyrightyear{2021}
\acmYear{2021}
\setcopyright{acmlicensed}\acmConference[AIES '21]{Proceedings of the 2021 AAAI/ACM Conference on AI, Ethics, and Society}{May 19--21, 2021}{Virtual Event, USA}
\acmBooktitle{Proceedings of the 2021 AAAI/ACM Conference on AI, Ethics, and Society (AIES '21), May 19--21, 2021, Virtual Event, USA}
\acmPrice{15.00}
\acmDOI{10.1145/3461702.3462521}
\acmISBN{978-1-4503-8473-5/21/05}

\begin{document}

\title{Fair Machine Learning Under Partial Compliance}

\author{Jessica Dai}
\affiliation{%
  \institution{Brown University}
  \city{Providence}
  \state{Rhode Island}
  \country{USA}
}
\email{jessica.dai@alumni.brown.edu}

\author{Sina Fazelpour}
\affiliation{%
  \institution{Northeastern University}
  \city{Boston}
  \state{Massachusetts}
  \country{USA}
}
\email{s.fazel-pour@northeastern.edu}

\author{Zachary C. Lipton}
\affiliation{%
  \institution{Carnegie Mellon University}
  \city{Pittsburgh}
  \state{Pennsylvania}
  \country{USA}
}
\email{zlipton@cmu.edu}


\begin{abstract}
Typically, fair machine learning research 
focuses on a single decision maker
and assumes that the underlying population is stationary.
However, many of the critical domains motivating this work 
are characterized by competitive marketplaces 
with many decision makers. 
Realistically, we might expect only a subset of them
to adopt any non-compulsory fairness-conscious policy, 
a situation that political philosophers 
call \emph{partial compliance}.
This possibility raises important questions:
how does partial compliance and the consequent 
strategic behavior of decision subjects
affect the allocation outcomes?
If $k$\% of employers were to 
voluntarily adopt a fairness-promoting intervention, 
should we expect $k$\% progress (in aggregate) 
towards the benefits of universal adoption,
or will the dynamics of partial compliance 
wash out the hoped-for benefits? 
How might adopting a global (versus local) perspective 
impact the conclusions of an auditor? 
In this paper, we propose 
a simple model of an employment market, 
leveraging simulation as a tool 
to explore the impact of both interaction effects
and incentive effects on outcomes and auditing metrics.
Our key findings are that at equilibrium:
\textbf{(1)} partial compliance by $k\%$ of employers
can result in far less than proportional ($k\%$) progress
towards the full compliance outcomes;
\textbf{(2)} the gap is more severe when fair employers 
match global (vs local) statistics;
\textbf{(3)} choices of local vs global statistics
can paint dramatically different pictures 
of the performance vis-a-vis fairness desiderata
of compliant versus non-compliant employers;
and \textbf{(4)} partial compliance 
based on local parity measures 
can induce extreme segregation.
Finally, we discuss implications for auditors 
and insights concerning the design 
of regulatory frameworks. 
\end{abstract}



\maketitle

\section{Introduction}
\label{sec:intro}
Responsible implementation of any allocation policy 
requires robust foresight about its likely impacts. 
In order to be useful,
such an analysis needs to take into account 
existing and emerging inter-dependencies 
between the policy and environmental factors
that shape the policy's long-term, situated consequences 
\citep{elster1992local, hansson2013ethics}. 
However, to date, most studies 
of the performance and bias 
of algorithms applied to allocation decisions
examine the algorithm in isolation,
ignoring the wider deployment context.
As a result, these analyses 
risk distorting our understanding 
of the impacts of specific algorithms,
and limit our ability to anticipate broader 
societal implications of algorithmic decision-making. 

Recently, a more critical thread 
in algorithmic fairness scholarship 
has called for a broader, 
systems-level approach to ``fairness'',
recognizing that algorithmic decisions 
do not happen in a vacuum
\citep{hu2018short, hardt2016strategic,liu2019delayed, Selbst2019,fazelpour2020algorithmic, green2018myth, kasy2020fairness}.
Decisions may have long-term ramifications for individual welfare 
beyond the snapshot captured at the time of prediction 
\citep{liu2019delayed, d2020fairness}.
Thus, shifting attention towards the agency, 
impacts, and responsibility 
of decision makers in context is imperative.

In this paper, we adopt such a systems-level approach 
to explore the setting where multiple decision makers
interact in a single labor market.
Rather than considering the fairness of policies
that a single decision maker might choose 
(i.e., the fairness of a single algorithm), 
we assume that there are several decision makers, 
whose decisions impact each another via market dynamics. 
While there are many possible settings 
in which a multi-decision maker scenario could take place---the 
provision of loans, for instance---we 
use the job market as a toy model for this scenario, 
both for simplicity and to set our work in dialogue
with the broader labor economics literature
addressing discrimination and partial compliance. 

Two factors complicate the situation.
First, employers vary in terms of their hiring policies,
especially concerning their adherence 
to fairness-promoting measures.
This situation of \emph{partial compliance} 
reflects the current reality of
predictive algorithms in hiring, 
which is characterized by heterogeneity across vendors
regarding the type of measures, if any, 
enforced for counteracting bias \citep{sanchez2020does}.
Second, complicating matters further, 
differences in hiring policies across institutions
can \emph{incentivize} strategic applications,
altering the distribution of candidates 
subsequently seen by employers \citep{elster1992local}.

We investigate these dynamics using simulation tools.
Our models consist of two types of agents:
applicants and employers.
The applicants each have a single ``score"
reflecting their perceived skill levels, 
and belong to one of two demographic groups:
one which has been historically disadvantaged, 
and associated with lower scores on average,
and one which has been historically advantaged, 
which has higher scores on average.
In this work, we take no position 
on the extent to which this disparity
is the result of systematic biases 
in the appraisal of the disadvantaged group,
or is an accurate reflection of skills 
that vary because of upstream discrimination in society.
Our general observations apply in both cases. 

The employers may either be 
fairness-conscious (\textit{compliant})---taking 
into account considerations of demographic parity 
\citep{Calders2009,feldman2015certifying}, 
or fairness-agnostic (\textit{non-compliant})---deciding 
solely on the basis of scores. 
We also explore settings where applicants decide 
the type of institutions to which they apply strategically 
in light of the different incentives 
afforded by these selection policies. 


We emphasize that our model is not intended 
as a realistic depiction of the labor market.
We do not claim to offer direct policy prescriptions. 
Instead, our purpose is to propose 
the simplest conceivable model 
that captures the effects of partial compliance.
By elucidating some basic qualitative insights 
regarding the impacts of partial compliance in allocative decisions,
we aim to clarify the associated set of concerns 
that must be accounted for by any regulator.
We argue that if even the most simple models 
evidence the complex interactive effects 
introduced by partial compliance behavior, 
then these effects must be 
considered when discussing the impact
of specific policies or algorithmic approaches.

\paragraph{Our findings}
Even with the simplest of assumptions, 
the relationships between 
the number of compliant institutions 
and various relevant metrics 
exhibit interesting phenomena:
\begin{enumerate}
\item Partial compliance
(by $k\%$ of employers)
can result in far less 
than proportional ($k\%$) progress
towards the full compliance outcomes.
\item This gap (between the benefit of $k\%$ compliance and $k\%$ of the benefit of full compliance)
is wider 
when compliant employers 
enforce demographic parity 
to match global (vs local) statistics.
\item Choices of 
global vs local statistics
can paint dramatically different pictures 
of the performance
of compliant (versus non-compliant) employers 
with respect to fairness considerations.
\item When coupled with incentive effects, 
partial compliance can induce 
extreme segregation across institutions.
\end{enumerate}
Our results illuminate a critical shortcoming
in current approaches to understanding fairness
in algorithmic-based allocations, 
and have significant implications 
for how we think about auditing decision makers
and assessing the potential benefits of regulation.
For example, simulations with our model show 
that even if a large fraction of employers 
voluntarily comply with a fairness-promoting policy, 
that does not necessarily mean 
that a  commensurate fraction of the benefit 
(relative to universal adoption)
has been realized. 
Consequently, a regulator assessing the urgency
of implementing fairness measure should take into 
account that even if only 20\% of the population
are non-compliant with a particular voluntary measure, 
they may be obstructing
a much larger share, say 50\% 
of the possible benefits of the policy.
Moreover, our findings suggest 
that in order to understand an employer's performance 
vis-a-vis fairness desiderata, 
it is not enough to look at statistics calculated 
based on the stream of candidates 
that apply to them---we
must also consider the way 
that the set of applicants that they encounter
may diverge from the demographics of the general population, 
and how these dynamics involve 
both interactions among the employers 
and strategic behavior among applicants. 

The rest of this paper is organized as follows:
In Section \ref{sec:related}, we survey literature 
from philosophy, (labor) economics, 
and the fair machine learning community, 
making connections to other work showing 
that the (partial) compliance 
among multiple decision makers
is an essential consideration
for assessing both moral responsibility 
and implementing practical measures.
In Section \ref{sec:simulation}, 
we introduce our model, 
including the parameters to our simulation
and several axes of variation that we explore.
In Section \ref{sec:results}, 
we discuss our experiments
and key results from those experiments.
Finally, Section \ref{sec:discussion}
provides a more critical discussion,
including implications for regulating
machine learning in allocative settings.

\section{Related Work}
\label{sec:related}
This work builds on several lines of research
in economics, fair machine learning, 
political philosophy, 
and computational social science. 
An extensive literature in economics 
models discrimination in employment.
\citet{becker1957economics} introduced the notion 
of \textit{taste-based discrimination}, 
where employers' distaste for hiring employees 
from a certain group results in them behaving
as though hiring a worker from the marginalized group
was associated with a higher cost (a ``disutility''),
despite workers from both groups 
being identical in terms of true skill level.
Becker also shows that this differential treatment 
among employers induces a sorting of minority employees
into the least discriminatory employers, 
with the equilibrium wage determined 
by the disutility 
of
the marginal discriminator. 
While our setup and motivation 
differ from Becker's,
with employers intervening to mitigate 
(rather than instigate) disparities,
this segregation effect induced 
by differential treatment across employers 
also appears in our model.

\citet{arrow1973theory} famously criticized Becker's model, 
arguing that discrimination thus characterized
would decrease competitiveness 
and be driven out of the market,
suggesting instead to focus on models 
of discrimination driven by imperfect information.
Along these lines, \citet{phelps1972statistical} introduced
a \textit{statistical} model for discrimination in hiring,
whereby disparities emerge 
due to differences in the difficulty
of measuring the true skill level 
of each group of employees. 
\citet{aigner1977statistical} build on this idea, 
arguing that economic discrimination ought to be measured 
by differential treatment 
conditioning on true skill.
By contrast, we take no position on whether 
observed scores accurately reflect
the employee's true skill level. 
Finally, \citet{coate1993will} 
address the long-term efficacy 
of affirmative-action policies, 
finding that, depending on 
specific parameter settings
in their model,
affirmative action can either eliminate stereotypes,
or appear to confirm (untrue) negative stereotypes.
As our ``fairness intervened" models 
are functionally affirmative-action policies, 
we also explore the long-term dynamics of such policies.
By contrast, we focus on the impacts  
of many employers adopting different policies 
on binary hiring decisions,
not on concerns regarding stereotypes or wages.

Another related line of work 
calls for more realistic assumptions 
about the social context of allocation 
\citep{hu2018short, liu2019delayed, fazelpour2020algorithmic, Selbst2019}. 
In the fair machine learning literature,
\citet{hu2018short}
called attention to dynamics
of employer-employee interactions,
modeling the labor market
as a series of principal-agent interactions.
They draw upon the same threads of the economics literature,
but focus on reputation and effort exertion.
\citet{liu2019delayed} focuses on credit ratings,
showing that with a simple but reasonable 
set of assumed dynamics,
certain fairness interventions 
can harm the very groups they are intended to protect.
\citet{hardt2016strategic, milli2018social, hu2018disparate, kleinberg2019classifiers, kilbertus2020fair}
all focus on the strategic behavior 
of individuals subject to automated decisions.
\citet{hu2020fair} consider fairness 
in a setting where multiple classifiers 
interact with one another in the same system.
Finally, 
\citet{rambachan2020economic}
approaches fair machine learning 
from an economic perspective,
constructing a social welfare function 
for a policymaker
and a private objective function 
for an algorithm designer, 
investigating the relationship 
between disclosure and regulation. 
While these works
recognize the problem
of framing decisions as classifications,
none focus on partial compliance,
the central issue in this paper.

By contrast, we focus on 
two aspects of deployment dynamics 
that, though critical in shaping 
the ethical impact of algorithms in context, 
tend to be abstracted away 
in standard evaluations of algorithmic systems. 
First, our model represents 
potential differences among decision-makers 
in adherence to ethical or legal obligations, 
thus relaxing the assumption of
a central decision-maker 
(or, equivalently, of full compliance), 
according to which all relevant agents 
comply with what justice demands of them.
Present in many philosophical theories of justice 
and implicitly assumed by many works in fair machine learning 
\citep{fazelpour2020algorithmic}, 
the full compliance assumption enables one 
to focus theorizing on the obligations 
that are the ``fair share'' of any agent. 
Nonetheless, recent philosophical works 
have cast doubt on whether theories 
developed under this assumption 
can provide sufficient practical guidance 
for agents in the actual world 
characterized by partial compliance 
\citep{Appiah2017,Valentini2012}.
This line of work considers when and how 
in circumstances of partial compliance 
agents might face obligations that differ 
from what would have been their fair share, 
had others complied
\citep{Valentini2012,Miller2011-MILTUT-3,schapiro2003compliance}. 
In the related labor economics literature, 
papers tend to focus on
determining the incentive structures
that promote or impede compliance with regulations 
such as minimum wage laws 
\citep{ashenfelter1979compliance, squire1997impact}, 
examining their macro-level impacts on the treatment 
of ``non-favored'' groups \citep{chang2007wage}.

Second, in our models, decision subjects are represented 
as agents capable of responding strategically 
to the incentive structure of the environment. 
While abstracted away in most analyses of algorithmic reliability,
this type of secondary effect is widespread 
in real-world allocation settings, 
and achieving foresight about its impacts 
is a priority for policy makers 
\citep{elster1992local, rambachan2020economic}. 
Our work contributes to 
emerging efforts 
in the fair machine learning literature 
towards broadening the scope of analysis 
to include these effects 
\citep{liu2020disparate,hardt2016strategic,d2020fairness}.
Moreover, in exploring the impact of these dynamics,
our work goes beyond assessments 
of algorithmic performance in static settings,
furthering research 
on the long-term impact of proposed interventions 
\citep{hu2018short,heidari2019long,liu2019delayed}.

We also build on recent research using 
simulation models to study fairness 
in ML systems \citep{d2020fairness}. 
While comparatively new in fair machine learning,
simulation studies represent a core methodology
in economics and sociology 
\citep{bianchi2015agent,centola2018behavior}, 
and are increasingly used by philosophers 
to study social dynamics in general \citep{zollman2013network}
and fairness in particular \citep{OConnor2019a,muldoon2016social}. 
Simulations are favored in these domains
owing to their ability to model emergent outcomes
of multiple interdependent decisions 
in non-stationary settings. 
Furthermore, particularly in the presence 
of heterogeneity in individual characteristics, 
simulations can yield insights
that are not readily available 
in traditional aggregate models, 
such as those
based on closed form solutions 
and/or systems of differential equations
\citep{kiesling2012agent}. 

\section{Experimental Setup}
\label{sec:simulation}
We now provide a detailed description of 
the models explored in our simulations and motivate their design.
In all of our models of a job market with partial compliance, 
all applicants have exactly two attributes: 
(i) a score, representing perceived skill for the job;
and (ii) a group identity. 
Applicants may belong either to 
the advantaged group with higher mean score (Group A)
or the marginalized group with lower mean score (Group B).
Across our experiments,
we consider two levels of representation 
in the broader population:
one where the disadvantaged group 
constitutes 25\% of the populations
and another where they constitute 50\%.
Our market contains a number 
of employers ($n=50$), 
$k\%$ of which may be compliant, and 
$(100-k)\%$ of which are non-compliant.

At each time step, some number of new applicants 
$(a = 1250)$
enter the job market.
Each newcomer to the applicant pool 
is randomly assigned a group membership
(according to population demographics).
Each applicant's score is drawn
from a normal distribution:
$\mathcal{N}(0, 1)$ for Group A, and 
$\mathcal{N}(-0.3, 1)$ for Group B.
Then, every applicant chooses one employer to apply to, 
and each institution hires $h=10$
applicants.
Once hired, applicants are removed from the market.
Additionally, we remove applicants 
that have not been hired after $10$ rounds. 

\subsection{How do institutions choose applicants?}
We consider three possible policies 
that institutions may adopt 
when choosing applicants to hire: 
one generic non-compliant strategy, 
and two possible fairness-conscious (i.e. ``compliant'') strategies, which satisfy some version of demographic parity.
\begin{enumerate}
    \item \textit{Generic policy}. 
    Non-compliant employers simply hire
    the $h$ highest-scoring applicants.
    \item \textit{Global parity policy}. 
    Compliant employers with the global parity policy 
    satisfy demographic parity with their hires, 
    with respect to global demographics; 
    this may or may not be the same as 
    the demographics of their applicant pool. 
    For example, if 25\% of the population belonged to Group B, 
    even if they accounted for 35\% 
    of applicants to a global-parity employer, 
    they would only account for 25\% of their hires.
    \item \textit{Local parity policy}. 
    Compliant employers with the local parity policy satisfy demographic parity 
    with respect to the demographics 
    of their applicant pool at that round; 
    in most cases, this is not the same as 
    the overall demographics of the environment. 
    For example, if 15\% of applicants 
    to a local-parity employer were from Group B, 
    then 15\% of the employer's hires will be from Group B, 
    even if Group B comprises 25\% of the entire population. 

\end{enumerate}

The latter two parity strategies are probabilistic---hiring 
$x\%$ from Group B \textit{in expectation}---rather 
than deterministically hiring a specific number from Group B 
based on a rounded proportion of available headcount. 
For simplicity, we only consider scenarios 
in which all compliant employers 
adopt the same strategy 
(either local or global).

\vspace{15px}
\noindent \textbf{Comments on demographic parity \quad}
Our operationalization of fairness 
in terms of demographic parity is not intended 
as an endorsement of this measure as the 
appropriate fairness measure in hiring settings. 
Rather, our choice is based 
on the widespread use of the measure 
in current practice \citep{sanchez2020does}, 
perhaps due to a perceived connection 
between the quantitative measure 
and disparate impact doctrine 
in the United States 
\cite{feldman2015certifying} 
and indirect discrimination regulations 
in the European Union \cite{sep-discrimination}\footnote{See \citet{Lipton2018a} and \citet{wachter2020fairness} for critical perspectives on the connection.}.

Additionally, 
if available scores 
accurately reflect ``true'' skill level, 
then the generic, non-compliant policy 
may actually be fair
according to 
to some proposed definitions of fairness,
such as calibration \citep{pleiss2017fairness}.
While our results are relevant regardless of 
the relationship between ``true'' and available 
scores, making this assumption 
means that our work can be also be re-interpreted as 
investigating the scenario where many intentionally-
compliant employers have different \textit{interpretations} 
of compliance---that is, employers are satisfying 
different definitions of fairness.

In the case that available scores \textit{do not} 
accurately reflect ``true'' skill level for Group B, 
consider a setting where the true skill distributions 
are identical, and the compliant policy involves 
correction for the score difference rather than setting 
explicit headcounts. More concretely, the score-correcting 
compliant policy will simply add the known difference 
in scores to the scores of all applicants from Group B, 
then hire the $h$ applicants with the highest (corrected) 
scores regardless of group membership---operationalizing 
fairness as treating individuals with the same 
true skill equally.
As it turns out, this setting yields identical results to the 
local parity policy: for any given set of applications, 
a local parity employer will hire the top $x\%$ of applicants
from each group. Meanwhile, a score-correcting employer 
corrects the scores of Group B, so that both groups have the
exact same score distributions. Then, hiring the top $x\%$ 
based on corrected scores is equivalent to hiring the top
$x\%$ from each group. 
However, we note that the two may diverge
when applicants' strategic behavior can be score aware. 


Finally, 
we note 
that both possible compliant
policies---local and global---are
constrained by the demographics of the 
applicant pool, even in the global parity case: 
for example, $25\%$ of Group B 
among all \textit{applicants}
may still reflect under-representation 
with respect to the entire population, 
which means that even a ``global parity''
employer only satisfies demographic parity 
with respect to the overall applicant pool, 
rather than the true global
population demographics.
\begin{figure*}[ht!]
    \centering
    \begin{subfigure}[b]{0.32\textwidth}
        \centering
		\includegraphics[width=1\textwidth]{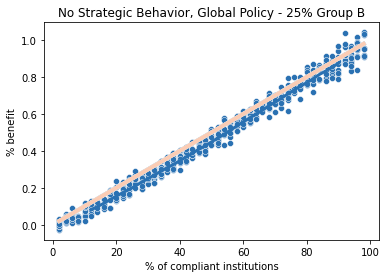}
    \end{subfigure}%
    ~
    \begin{subfigure}[b]{0.32\textwidth}
        \centering
        \includegraphics[width=1\textwidth]{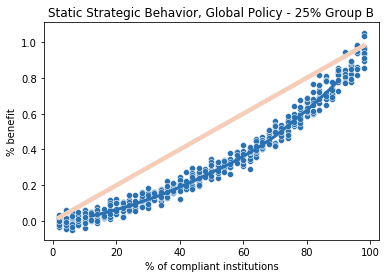}
    \end{subfigure}
    ~
    \begin{subfigure}[b]{0.32\textwidth}
        \centering
        \includegraphics[width=1\textwidth]{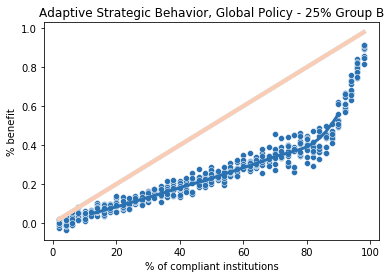}
    \end{subfigure}
    \\
    \begin{subfigure}[b]{0.32\textwidth}
         \centering
         \includegraphics[width=1\textwidth]{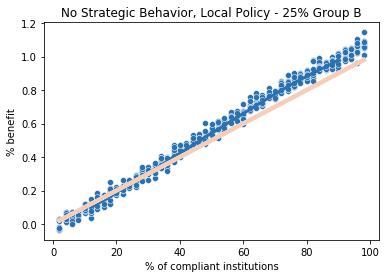}
     \end{subfigure}
    ~
    \begin{subfigure}[b]{0.32\textwidth}
        \centering
		\includegraphics[width=1\textwidth]{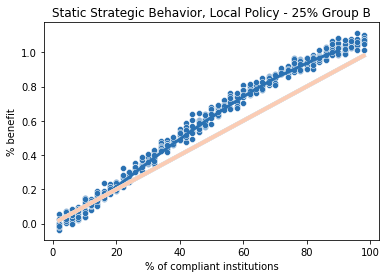}
    \end{subfigure}%
    ~
    \begin{subfigure}[b]{0.32\textwidth}
        \centering
		\includegraphics[width=1\textwidth]{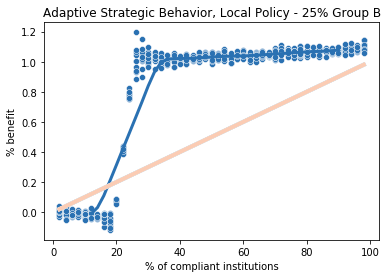}
    \end{subfigure}%
    \caption{Benefit as measured by demographic parity for different 
    institution policies. 
    Far left plots show market where all applicants 
    pick employers uniformly at random; 
    as expected, we see exactly linear gain. 
    In the center column, applicants have 
    a slight preference for 
    a more favorable employer
    (compliant for Group B, non-compliant for group A), 
    and in the far right plots, 
    applicants have an adaptive strategy. 
    }
    \label{fig:dempar_static}
\end{figure*}
    \subsection{How do applicants choose institutions?}
We also consider three possible strategies 
that applicants may employ
when choosing institutions to apply to. 
Let 
$p_{G \in A, B}$
represent 
the probability of an applicant 
from group G (either A or B) 
choosing to apply to a \textit{compliant} institution,
scaled by the total number of compliant institutions.
Like the employer policies, 
these strategies are stochastic.

\begin{enumerate}
    \item \textit{Completely at random}. 
    All applicants from both groups are equally likely 
    to apply to institutions of either type; 
    hence, 
    $p_A = p_B = k$.
    This reflects \textit{no} strategic behavior, 
    i.e., applicants have no sensitivity to incentives.
    \item \textit{Static preference}. 
    Over the course of the simulation, 
    all applicants from Group A have a fixed preference 
    for applying to a non-compliant institution, 
    and all applicants from Group B have a fixed preference 
    for applying to a compliant institution; 
    hence,
    $p_A < k < p_B$.
    This reflects strategic behavior,
    where applicants have some knowledge 
    about the 
    nature of each institution's policies, 
    but no access to additional information 
    over the course of the simulation---that is, 
    applicants are sensitive to incentives 
    but have limited knowledge of the system. 
    \item \textit{Dynamic preference}. 
    Over the course of the simulation, 
    $p_A$ and $p_B$
    are adjusted for each round 
    based on the results 
    of the previous round. 
    For each group, if that group's acceptance rate 
    in \textit{compliant} institutions is greater than 
    its acceptance rate in \textit{non-compliant} institutions,
    then the log odds ratio 
    $\ln({p_G}/({1-p_G}))$
    is increased by a constant amount 
    $0.05$.
    Otherwise, it is decreased by the same amount. 
    Equilibrium for each group is reached 
    when the probability of being accepted at a parity institution 
    is the same as the probability 
    of being accepted at a generic institution.
    This reflects strategic behavior 
    where applicants are aware 
    of their group membership, 
    have access to new information at each timestep, 
    and are able to update their strategy accordingly. 
\end{enumerate} 


\vspace{15px}
\noindent \textbf{Comments on applicant strategy and agent-based modeling \quad}
While we do not claim that these strategies 
exactly model the decision making processes 
of individuals in the real world, 
these coarse approximations of aggregate behavior
yield valuable qualitative insights. 
Though we use a simple toy model, 
the core motivations for its design 
are grounded in reality.
The hiring platform Applied, for example, 
claims that fairness-conscious hiring policies 
result in increased applications 
from minority groups \citep{applied}. 
It is impossible to exactly quantify 
the degree to which either applicant strategy (static or dynamic) 
represents ``true'' behavior. 
However, as mentioned in Section \ref{sec:related}, 
simulation studies are a core 
methodology in both economics and philosophy, 
and in this work, the value of simulation 
is to test the qualitative impact 
of some sort of applicant strategy. 

\section{Results}
\label{sec:results}


In all of our experiments, 
we vary the number of compliant institutions 
(out of $50$ total) from $0$ to $50$. 
For each number of compliant institutions, 
we run ten trials of the simulation.
For each trial, we run the simulation 
until it reaches equilibrium:
$100$ steps for static applicant strategy,
and $200$ steps for adaptive applicant strategy.
We then continue running the simulation 
for the same number of additional timesteps
and calculate statistics from each trial
based on the post-equilibrium timesteps. 
In all of our plots, one dot reflects 
the statistics calculated from a single trial. 

\begin{figure}[ht!]
    \centering
    \begin{subfigure}[t]{0.24\textwidth}
        \centering
		\includegraphics[width=1\textwidth]{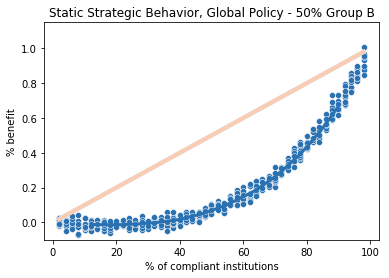}
    \end{subfigure}%
    ~ 
    \begin{subfigure}[t]{0.24\textwidth}
        \centering
        \includegraphics[width=1\textwidth]{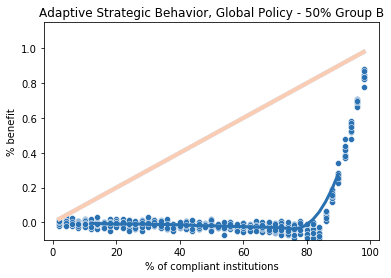}
    \end{subfigure}

    \begin{subfigure}[t]{0.24\textwidth}
        \centering
		\includegraphics[width=1\textwidth]{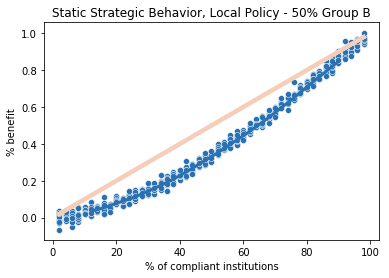}
    \end{subfigure}%
    ~ 
    \begin{subfigure}[t]{0.24\textwidth}
        \centering
        \includegraphics[width=1\textwidth]{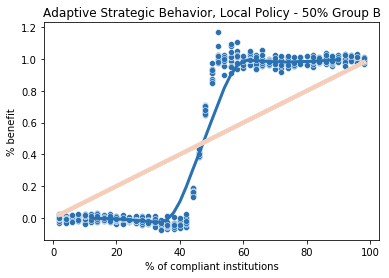}
    \end{subfigure}
    \caption{Aggregate statistics when Group B is 50\% of the population; 
    benefit is measured by overall demographic parity. 
    Left: static applicant strategy; right: adaptive applicant strategy.}
    \label{fig:50b}
\end{figure}

\vspace{15px}
\noindent \textbf{Sublinear gain \quad}
Our first key finding is that when employees apply strategically, 
then under partial compliance, the aggregate benefit 
from an additional compliant employer depends strongly 
on how many institutions are already compliant. 
In Figure \ref{fig:dempar_static}, 
all employees apply 
with the strategy of static preference: 
that is, knowing that compliant employers 
are more likely to hire Group B applicants, 
and that non-compliant employers are 
more likely to hire Group A applicants, 
employees from Group B apply 
to compliant employers with probability $0.55$ 
(scaled by number of each type of employer) 
and employees from group A apply
to non-compliant employers with probability $0.55$. 
The y-axis is scaled demographic parity, 
where $y=0$ corresponds to the 
disparate impact score 
$\frac{P(\textit{hired} | B)}{P(\textit{hired} | A)}$ 
when all employers are non-compliant 
(with our main experimental parameters, 
this is approximately 0.75),
and $y=1$ corresponds to ``perfect" parity. 
One might hope that $k\%$ compliance would correspond 
to at least $k\%$ of the benefits, 
a condition that we denote \textit{linear gain}.
In Figures \ref{fig:dempar_static} and \ref{fig:50b}, 
this is illustrated by the light peach line.  

Notably, when all compliant institutions satisfy 
fairness with respect to \textit{global} statistics,
the partial compliance curve is convex, 
illustrating \textit{sub}linear gain---$k\%$ compliance 
always gives less than $k\%$ of the attainable benefit.
Perhaps this should not be a surprising result.
The baseline demographic parity (\% 
benefit = 0, at 0\% compliance) reflects a scenario where 
each (non-compliant) employer receives an applicant pool 
that reflects the overall demographics of the system
(i.e., if 25\% of all applicants in the system are from 
Group B, then on average 25\% of non-compliant employers'
applicants also are from Group B). 
In order for \textit{linear
gain} to occur, 
then at $k\%$ compliance, 
all $(100-k)\%$ non-compliant employers 
must hire at the same rate 
as they were at $0\%$ compliance 
even as the $k\%$ compliant employers
hire exactly in accordance 
with global demographic parity. 
However, due to applicant strategy, 
the distribution of applicants 
to non-compliant employers 
at $k\%$ compliance
no longer reflects the demographics 
of the system.
Instead, non-compliant employers 
see relatively more Group A applicants 
and relatively fewer Group B applicants. 
As a result, the non-compliant hiring strategy results in 
an even lower percentage of Group B
(as a proportion of overall hires) 
than at $0\%$ compliance,
giving rise to sublinear gain. 

Under local parity policies, the partial compliance curve
can actually reflect \textit{super}linear gain, 
as when Group B constitutes 25\% of the population. 
However, when Group B constitutes 50\% 
of the population (Figure \ref{fig:50b}), these dynamics change: 
local parity policies now also induce \textit{sub}linear gain,
and the global parity curve indicates 
a more pronounced sublinear gain.

Regardless of whether Group B comprises 25\%
or 50\% of the population, 
following the \textit{global} parity policy 
leads to comparatively worse gains 
than following the \textit{local} parity policy---that is, 
for any given $k\%$ compliant institutions, 
the percent benefit when employers 
satisfy global parity 
is lower than when employers satisfy local parity. 
The explanation, both for super/sub-linearity of local parity
policies, and for 
why sublinear gain under global parity is always
worse than under local parity, 
lies in the flexibility 
that a local parity policy affords. 
Under global parity policies, 
the fraction of hires 
that a compliant institution 
can make from Group B 
is 
fixed based on their share of the underlying population.
However, with local parity policy, it is possible for
all $k\%$ of the 
compliant employers to 
allocate their entire headcount 
to Group B 
(in the event that Group B comes 
to represent 100\% of their applicants).
Thus, under local parity, 
compliant employers 
are able to take on 
more than their ``fair share''
(to borrow terminology from 
the philosophy literature on partial compliance).

\vspace{15px}
\noindent \textbf{Static vs adaptive applicant strategy \quad}
When employees are able to update their 
application strategy at each timestep, 
interesting dynamics emerge 
(Figure \ref{fig:dempar_static}, \ref{fig:50b}). 
Recall that the likelihood of employees 
from a given group applying to 
each type of employer (compliant vs non-compliant) 
is adjusted based on group-wise acceptance rates 
from the previous timestep.
Hence, equilibrium for each group is reached 
when that group encounters the same acceptance rate
from both compliant and non-compliant employers. 
Under global parity policies, 
the first 80\% of compliant institutions 
are only able to push the macro-level statistics 
around \textit{halfway} to parity; 
the remaining 50\% of benefits relies entirely 
on the last 20\% of employers 
becoming
compliant. 
Interestingly, under local parity policies
when Group B is 25\% of the population,
the first 20\% of compliant employers 
have functionally no effect on the macro-level view of fairness,
while complete parity is achieved
by the time around 30\% of employers are compliant. 
\begin{figure}[ht!]
    \begin{subfigure}[t]{0.5\linewidth}
        \centering
		\includegraphics[width=\textwidth]{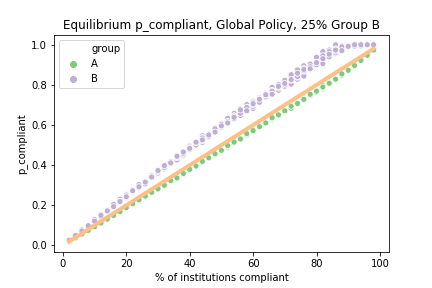}
    \end{subfigure}%
    \begin{subfigure}[t]{0.5\linewidth}
        \centering
        \includegraphics[width=\textwidth]{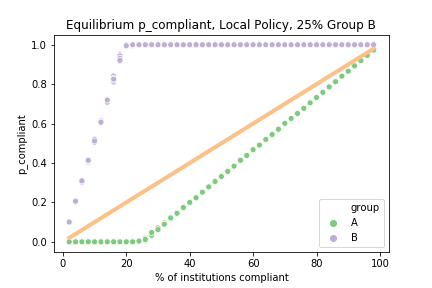}
    \end{subfigure}
    \caption{Groupwise \textit{equilibrium} probability 
    ($p_A$ and $p_B$ described in Section \ref{sec:simulation})
    of applying to either compliant or non-compliant employers, under adaptive applicant strategy.
    The orange line indicates the 
    $p$
    reflecting no preference
    (i.e. probability determined 
    solely by the 
    proportion
    of compliant institutions 
    currently in the system,
    $p = k$). 
    Left: global parity policy; 
    right: local parity policy.}
    \label{fig:adapt_ps}
\end{figure}
\begin{figure*}[t]
    \centering
    \begin{subfigure}[b]{0.49\linewidth}
        \centering
		\includegraphics[width=\textwidth]{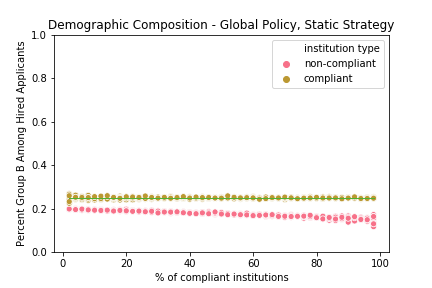}
    \end{subfigure}%
    ~ 
    \begin{subfigure}[b]{0.49\linewidth}
        \centering
		\includegraphics[width=\textwidth]{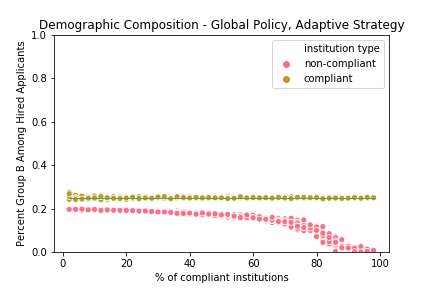}
    \end{subfigure}%
    \\
    \begin{subfigure}[b]{0.49\linewidth}
        \centering
		\includegraphics[width=\textwidth]{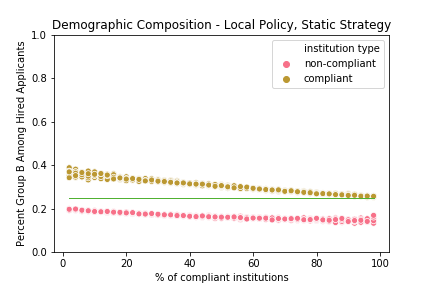}
    \end{subfigure}%
    ~
    \begin{subfigure}[b]{0.49\linewidth}
        \centering
		\includegraphics[width=\textwidth]{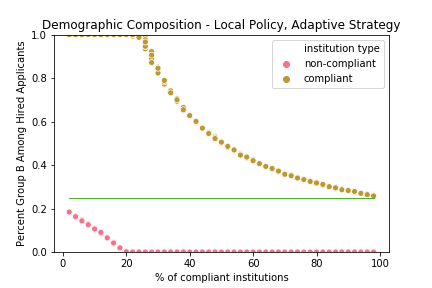}
    \end{subfigure}%
    \caption{
    Demographic composition among hired employees, 
    by institution type---in these graphs, 
    Group B is 25\% of the population. 
    Top row: Applicants employ a \textit{static} strategy.
    Bottom row: applicants employ an \textit{adaptive} strategy.
    Light green horizontal line indicates percentage of Group B in population.} 
    \label{fig:segregation}
\end{figure*}
For intuition as to why this is the case, 
we can look at the equilibrium probabilities 
for applying to either type of employer.
Figure \ref{fig:adapt_ps} shows 
that under local parity policies, 
the equilibrium 
$p_B$
(probability of applying to a compliant employer 
for Group B)
quickly goes to 1.
With 20\% or more compliant employers,
Group B always applies almost exclusively 
to compliant institutions.
Meanwhile, until 26\% or more employers are compliant,
Group A applies almost exclusively
to non-compliant institutions. 
Under global parity policies, 
the difference in preference induced by partial adoption 
of the fairness-promoting policy is less severe. 

\vspace{15px}
\noindent\textbf{The emergent demographic composition of institutions}
A closer look at institution-specific outcomes 
reveals that at equilibrium, 
strategic applications can result 
in \emph{homogeneity} within institutions 
and \emph{segregation} across institutions.
In the case of global parity policies, 
the dramatic increase in aggregate parity
(Figure \ref{fig:dempar_static}, right column) 
is coupled with a precipitous drop-off 
in the percentage of hired applicants 
belonging to Group B in non-compliant institutions
(Figure \ref{fig:segregation}, bottom left).
The situation is even more dire under local parity policies,
as the the equilibrium strategies mean 
that non-compliant institutions 
have \textit{no} hired applicants 
(or indeed, applications)
from members of Group B 
(Figure \ref{fig:segregation}, bottom right). 
Notably, though the aggregate parity curves 
under the global policy 
do not look so different 
in Figure \ref{fig:dempar_static},
the segregation effects do \textit{not} occur 
when applicants operate under 
a \textit{static} application strategy: 
while partial compliance has
some impact on the overall demographic composition of hired
employees, the percentage of Group B 
never approaches zero 
(Figure \ref{fig:segregation}, top row). 

\vspace{15px}
\noindent\textbf{The impact of the original demographic makeup
on adaptive applicant strategy \quad} 
%
When employees were applying to firms 
under a \textit{static} strategy,
the impact of changing from a scenario 
where Group B is 25\% of the population
(Figure \ref{fig:dempar_static})
to one where Group B is 50\% of the population 
(Figure \ref{fig:50b}),
while significant, affects aggregate statistics 
in similar ways at all levels of compliance 
and for both global and local parity policies.
However, when applicant strategies are adaptive,
increasing the proportion of Group B in the population
(Figure \ref{fig:50b})
means that under global parity policies, 
the first 80\% of compliant institutions---despite 
reaching 50\% of the benefit 
when Group B was 25\% of the population
(Figure \ref{fig:dempar_static})---actually 
have \textit{no} impact on aggregate demographic parity. 
The critical tipping point, however, 
remains the same, at 80\% compliance. 
Under local parity policies, on the other hand, 
the overall shape of the aggregate parity curve 
remains the same---two large regions
with either zero or perfect parity, 
and one small intermediary transition region---but 
when Group B comprises 50\% of the population,
the critical transition region is 
between 40\%-50\% compliance, 
rather than 20\%-30\% compliance.

\section{Discussion}
\label{sec:discussion}

Our simulations illustrate several fundamental 
but commonly overlooked issues 
that plague the ethical evaluation
and governance of algorithmic tools 
in consequential allocation settings. 
While our results do not imply 
specific or prescriptive policy solutions,
they do raise critical questions about 
the design and adoption of fair policies.


\vspace{15px}
\noindent\textbf{Beyond narrow assessments of fairness: diversity and integration \quad}
Consider first that,
in many allocative contexts,
task-related utility and fairness
are not the only desiderata. 
For example, in hiring contexts, 
diversity within the workforce
is intrinsically valuable,
both due to its potential 
to enhance team performance
and on moral and political grounds 
\citep{steel2019information,page2019diversity}. 
While recent work in fair ML 
has begun to consider the interaction 
between diversity, utility and fairness 
\citep{celis2016fair,drosou2017diversity}, 
most analyses remain restricted to static settings, 
focused on individual decision-makers,
neglecting the interactions among their decisions
and those of their peers
and the influence of dynamic factors,
such as incentive effects,
on long-term policy consequences.
Consider what \citet{steel2018multiple} refer to 
as the representative concept of diversity
(see also \citet{smith2017diversity}), 
motivated by concerns about democratic legitimacy,
which requires the distributional properties 
of the selected group to match 
those of the general population. 
The global demographic parity measure 
thus tracks this notion of diversity.
Viewed through a static lens, 
and setting aside the influence of incentives 
on the choice behavior of applicants, 
the same connection could be said to hold
between the diversity concept and 
local demographic parity measures. 
Indeed, this has led some authors 
to roughly equate these notions 
of diversity and fairness \citep{celis2016fair}.
The situation becomes more complicated, however, 
once the dynamics of adaptive application
are taken into account.
Here, the appearance of (ostensibly desirable) 
parity at the aggregate level 
conceals the detrimental impact 
of local parity policies on diversity
within the workforces of the individual employers.
These outcomes can emerge absent
any explicit desire for segregation
on the part of applicants or employers; 
rather, they are a consequence of the dynamics 
of incentive effects under partial compliance. 
In addition to stripping institutions 
of the benefits of diversity, 
the resulting segregation 
can exacerbate the homophily-based processes 
that, according to a number of authors 
\citep{Anderson2010,OConnor2019a}, 
can cultivate or amplify injustice.

\begin{figure}[t]
    \centering
    \includegraphics[width=0.5\textwidth]{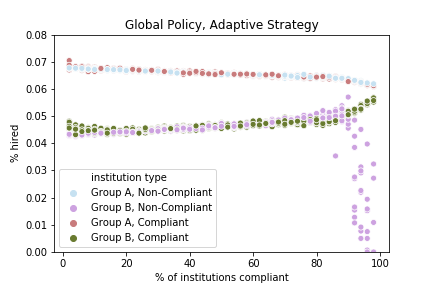}
    \caption{Percent hired per group per institution type. 
    Group B is 25\% of the overall population. 
    }
    \label{fig:global_hirerates}
\end{figure}

\vspace{15px}
\noindent \textbf{The aims and the value-alignment of regulation \quad}
The above discussion indicates the urgent need 
to clarify the aims and value orientation of regulation. 
As \citet{rambachan2020economic} 
note,
many discussions of regulation 
related to algorithmic fairness 
are fundamentally
concerned with selecting a policy 
that will generate an optimal 
distribution of outcomes.
Naturally, this requires first 
deciding what constitutes 
the optimal outcome distribution. 

It is useful to frame this issue 
by inquiring about the
aims of the policy that might 
support the enforcement
of local (vs global) demographic parity. 
In practice, 
demographic parity is popular, 
perhaps owing to the $80\%$ rule,
which is sometimes invoked as a statistical test 
in the first phase of disparate impact cases 
\citep{feldman2015certifying}. 
Note, however, that this connection 
does not provide a blind endorsement
of this form of parity as that 
which ought to be \emph{enforced}. 
Certainly, demographic parity can be 
a part of a \emph{diagnostic toolbox}, 
serving to indicate disparities 
that \emph{could}, but need not, 
indicate underlying discrimination
\citep{barocas2016big,lipton2018does}.
When precisely measured, demographic disparity 
can signal moral or legal failings 
with that particular employer
which lie outside the narrow scope 
of the quantitative measure itself. 
However, even when the disparity \emph{is} 
a symptom of underlying ethical troubles
with an allocation policy, 
enforcing the measure may be a misguided remedy 
to addressing these troubles  
(e.g., when the trouble lies with 
the choice of target outcomes or labels). 

Another way of motivating the enforcement 
of (some form of) demographic parity
is by reference to an employer's wish 
to implement \textit{affirmative action}. 
That is, employers may wish 
to enforce demographic parity,
and so preferentially select applicants 
on the basis of their group membership,
as a means of complying with a moral obligation 
to increase the representation 
of historically disadvantaged 
social groups in their institutions.
This interpretation resonates with the suggestions 
that, in some cases, the use of measures 
such as demographic parity is motivated 
by the ``long-term societal goal'' 
of living in a society where protected attributes 
are independent of task-relevant outcomes 
\citep{barocas-hardt-narayanan}.
However, specifying the relation 
between demographic parity and affirmative action
requires clarity about the underlying aim 
and justifications of the latter---issues 
that vary radically across different models 
of affirmative actions \cite{Anderson2010}---and 
considerations of whether the former 
indeed serves those aims.
Crucially, our results indicate 
that, even with minimal incorporation of deployment dynamics, 
the (partial) adoption of local demographic parity 
is inconsistent with prominent 
\emph{future-oriented} 
justifications of affirmative action.
In particular, the emergence of 
between-institute segregation and 
a lack of within-institute diversity in our simulations 
indicate that 
partial compliance with the measure 
can result in significant conflicts
with diversity-based \citep{sep-affirmative-action} 
and integration-based \citep{Anderson2010} 
arguments for affirmative action.

Of course, one could adopt a different model 
of affirmative action 
to motivate the enforcement of demographic parity. 
For instance, depending on 
the interpretation of scores in our model 
(e.g., as a result of past, upstream injustices,
or as an outcome of ongoing biases
in an employer's hiring practices),
the measure could be connected 
to compensation-based
\citep{thomson1973preferential}
or discrimination-offsetting 
\citep{warren1977secondary} justifications. 
Each of these models faces its own set of challenges, 
including discordance with the actual practice of law,
failure to account for the \textit{weight} 
given to social categories in preferential selection,
engendering the expressive harm of \textit{stigmatization}, 
and undermining the societal legitimacy 
of affirmative action
\citep{Anderson2010, sep-affirmative-action}.

While adjudicating between 
different models of affirmative action 
is beyond the scope of this paper, 
it raises an important concern:
Decisions about the aims and the alignment 
of regulation are value-laden to their core. 
As a result, these decisions should be made transparently, 
and on the basis of an integrated consideration
of the relevant moral and political models. 
Importantly, our results show 
that individual efforts (or the lack thereof)
to promote fairness can remain out of sight
unless assessed through a more comprehensive, dynamic lens.
Analyses of the kind carried out in this paper
can not only bring these value judgments into the open,
but also complement theorizing about 
which moral and political models 
we should prefer, and why. 
For example, analyses of deployment dynamics 
can offer qualitative insights 
about other meaningful endpoints 
and value-relevant considerations (e.g., diversity) 
that are likely to be relevant 
to assessing the desirability 
of alternative policies in context. 
Such approaches can thus contribute 
to recent calls for enriching the evaluation 
of downstream impacts of algorithmic decision-making
\citep{parikh2019regulation}. 
Viewed from this perspective, 
far from simply being a burden to be neglected 
in the context of ethical design, 
assessment of deployment dynamics 
can guide our deliberations in such contexts.

\vspace{15px}
\noindent \textbf{Partial compliance and the design of appropriate auditing frameworks \quad}
The type of partial compliance explored in this paper
is a simple representation 
of the kinds of heterogeneity 
that exist in the adoption
of fairness-promoting measures 
among various employers 
both in the use of algorithmic tools in hiring \cite{sanchez2020does}
and in hiring more generally.
The varied choices of measures
is \textit{a consequence} of the 
ambiguity of current regulatory frameworks.
Indeed, the laxity of constraints provides 
even the \textit{non-compliant} employers 
in our simulations with a claim to fairness.
That is, insofar as these employers 
have access to the ``ground truths'' for skill scores,
they can be seen as employing a perfect predictor 
that satisfies a number of other fairness desiderata 
suggested in the literature, 
such as parities in sensitivity, 
specificity, and precision across groups 
\citep{dieterich2016compas,chouldechova2017fair}.
Similar to the evaluative practices 
that inform them, 
these regulatory frameworks 
appear to be based on unrealistic assumptions 
and abstractions of the problem domain. 

Our exploration of the dynamics of partial compliance 
raises central concerns 
that should inform judgments about 
both the need for regulation
and the form that it should take. 
The discussion above relies heavily on 
the assumption that a regulator 
would be able to bring about 
something approximating full compliance to begin with. 
Determining how the behavior 
of individual decision makers 
compares to the behavior of all decision makers---and by extension, 
whether partial compliance is occurring---is 
therefore a critical concern for
any regulatory regime.

Existing approaches to auditing have focused 
on examining the performance 
of a single algorithmic decision maker 
\citep{raji2020closing, green2019disparate, sandvig2014auditing}. 
Similarly, \citet{wilsonbuilding}'s work with Pymetrics, 
a hiring platform that uses local demographic parity, 
explicitly considers only the pool of applicants 
that Pymetrics receives, 
seeking to verify the extent to which the selection procedure 
adheres to this version of demographic parity.
\footnote{Our emphasis here is on the fact 
that the audit is solely focused on Pymetrics, 
not to claim that Pymetrics functions as a decision maker 
in the same way as employers do in our simulation. 
Additional discussion of the Pymetrics audit's coverage, 
while merited, is out of scope for this work.}

However, in addition to highlighting
the potential cost to diversity and integration, 
our analysis shows that fairness statistics
reported by each employer 
are impacted not only by their policies, 
but also by those of their competitors. 
In our simulation, 
at equilibrium under partial compliance, 
when employers adopt the global parity policy,
an auditor looking at the fractions of applicants from 
Group A and Group B hired might erroneously
conclude that compliant and non-compliant 
employers were behaving similarly
(Figure \ref{fig:global_hirerates}).
However, this mistaken view fails to account
for the incentive effects, 
whereby compliant employers
come to receive many more applications
from members of the disadvantaged group.
Thus, when auditing performance 
vis-a-vis ethical desiderata,
we may not be able to determine 
how a firm is performing 
without also evaluating their peers.

Phenomena of this sort are not 
exclusive to partial compliance settings.
\citet{d2020fairness}, which study the 
long-term impact of (fair) decisions,
find a similar instance of Simpson's paradox
where enforcing equal opportunity
at each point in time does not result in 
equal opportunity in the aggregate,
in the presence of interactive effects between decisions 
and the characteristics of decision subjects. 
Taken together, these results suggest a need
for auditors to investigate not only the 
distributions of outcomes given the data,
but the actual underlying policy.
To this end, \citet{rambachan2020economic}, 
who study the construction of ideal (fair)
policy from the perspective of a regulator,
find a particularly interesting result: 
the ideal disclosure regime is
one where individual decision makers
must disclose all information 
about their algorithm and decision rule, 
and the effectiveness of regulation 
is substantially diluted 
when disclosure is more limited.
Finally, although the downstream consequences 
of \textit{regulation} in a dynamic environment 
is beyond the scope of this work,
viewing regulation under a dynamic lens 
suggests that the potential 
for partial compliance to mask 
the efforts of compliant institutions 
may provide an incentive 
for those institutions 
to share information about their policies 
with auditors 
despite the desire to protect proprietary information, 
because it may help differentiate 
themselves from non-compliant institutions. 


In some sense, the abstractions in our model 
\emph{under}estimate the implications of partial compliance 
for current regulatory and evaluative practices. 
This is because our model represents partial compliance
only with respect to \emph{concurrent} policies 
in a competitive marketplace of hiring. 
That is, we do not consider
allocations that are \emph{upstream} (e.g., in education) 
and \emph{downstream} (e.g., promotion, mobility across work sector, banking) 
from hiring decisions, 
each made by decision-makers who may or may not 
adhere to their legal (or moral) obligations.

\citet{elster1992local} makes vivid 
the significance of such allocations 
for the well-being and opportunities of individuals:
\begin{quote}
    The life chances of the citizen in modern societies ... 
    depend on allocations made by relatively autonomous institutions,
    beginning with admission or nonadmission to nursery school 
    and ending with admission or nonadmission to nursing homes.
    One could write the fictional biography of a typical citizen, 
    to depict his life as shaped by successive encounters 
    with institutions that have the power 
    to accord or deny him the scarce goods 
    that he seeks \citep[p.~2]{elster1992local}.
\end{quote}
Despite the potential of unexpected outcome
due to robust couplings between policies 
at successive allocative settings,
the implications of partial compliance at successive stages
remain under-investigated by the fair ML community. 
This is a challenge that requires 
a concerted interdisciplinary effort by the community. 
Indeed, providing practical guidance under partial compliance 
poses a challenge to traditional frameworks
of distributive justice in political philosophy. 
While looking to these frameworks 
for robust conceptual underpinnings of fairness measures
can be fruitful \cite{binns2018fairness}, 
they were mainly concerned with modeling 
the re-distributive obligations 
of a nation state towards its citizens 
from the perspective of economic justice.
However, when our focus is to provide guidance 
to relatively autonomous decision-makers 
using ML tools in local allocative settings,
we can no longer simply operate with the same assumptions.
Responsible innovation in general \citep{grunwald2014technology} 
and ethical deployment of algorithmic-based decision-making
in particular \cite{fazelpour2020algorithmic} 
require more comprehensive foresight studies 
that are equipped to deal with 
the complexities of the deployment context. 
We hope that our work contributes 
a few preliminary steps towards this aim. 

\begin{acks}
Zachary Lipton thanks the Block Center for Technology and Society, Amazon AI, and NSF: Fair AI Award IIS2040929 for supporting ACMI lab's research on the responsible use of machine learning. ZL is also grateful to PwC USA for funding this research through the Digital Transformation and Innovation Center sponsored by PwC.
Sina Fazelpour thanks the Block Center for Technology and Society and the Social Sciences and Humanities Research Council of Canada (award number 756-2019-0289).
Jessica Dai was supported in part by a LINK award from Brown University.

\end{acks}

\bibliographystyle{ACM-Reference-Format}
\bibliography{refs}


\begin{thebibliography}{66}


\ifx \showCODEN    \undefined \def \showCODEN     #1{\unskip}     \fi
\ifx \showDOI      \undefined \def \showDOI       #1{#1}\fi
\ifx \showISBNx    \undefined \def \showISBNx     #1{\unskip}     \fi
\ifx \showISBNxiii \undefined \def \showISBNxiii  #1{\unskip}     \fi
\ifx \showISSN     \undefined \def \showISSN      #1{\unskip}     \fi
\ifx \showLCCN     \undefined \def \showLCCN      #1{\unskip}     \fi
\ifx \shownote     \undefined \def \shownote      #1{#1}          \fi
\ifx \showarticletitle \undefined \def \showarticletitle #1{#1}   \fi
\ifx \showURL      \undefined \def \showURL       {\relax}        \fi
\providecommand\bibfield[2]{#2}
\providecommand\bibinfo[2]{#2}
\providecommand\natexlab[1]{#1}
\providecommand\showeprint[2][]{arXiv:#2}

\bibitem[\protect\citeauthoryear{Aigner and Cain}{Aigner and Cain}{1977}]%
        {aigner1977statistical}
\bibfield{author}{\bibinfo{person}{Dennis~J Aigner} {and}
  \bibinfo{person}{Glen~G Cain}.} \bibinfo{year}{1977}\natexlab{}.
\newblock \showarticletitle{Statistical theories of discrimination in labor
  markets}.
\newblock \bibinfo{journal}{\emph{Ilr Review}} \bibinfo{volume}{30},
  \bibinfo{number}{2} (\bibinfo{year}{1977}), \bibinfo{pages}{175--187}.
\newblock


\bibitem[\protect\citeauthoryear{Altman}{Altman}{2020}]%
        {sep-discrimination}
\bibfield{author}{\bibinfo{person}{Andrew Altman}.}
  \bibinfo{year}{2020}\natexlab{}.
\newblock \showarticletitle{{Discrimination}}.
\newblock In \bibinfo{booktitle}{\emph{The {Stanford} Encyclopedia of
  Philosophy} (\bibinfo{edition}{summer 2020} ed.)},
  \bibfield{editor}{\bibinfo{person}{Edward~N. Zalta}} (Ed.).
  \bibinfo{publisher}{Metaphysics Research Lab, Stanford University}.
\newblock


\bibitem[\protect\citeauthoryear{Anderson}{Anderson}{2010}]%
        {Anderson2010}
\bibfield{author}{\bibinfo{person}{Elizabeth Anderson}.}
  \bibinfo{year}{2010}\natexlab{}.
\newblock \bibinfo{booktitle}{\emph{{The Imperative of Integration}}}.
\newblock \bibinfo{publisher}{Princeton University Press},
  \bibinfo{address}{Princeton}.
\newblock


\bibitem[\protect\citeauthoryear{Appiah}{Appiah}{2017}]%
        {Appiah2017}
\bibfield{author}{\bibinfo{person}{Kwame~Anthony Appiah}.}
  \bibinfo{year}{2017}\natexlab{}.
\newblock \bibinfo{booktitle}{\emph{{As If: Idealization and Ideals}}}.
\newblock \bibinfo{publisher}{Harvard University Press},
  \bibinfo{address}{Cambridge}.
\newblock
\showISBNx{9780674975002}


\bibitem[\protect\citeauthoryear{Arrow et~al\mbox{.}}{Arrow
  et~al\mbox{.}}{1973}]%
        {arrow1973theory}
\bibfield{author}{\bibinfo{person}{Kenneth Arrow} {et~al\mbox{.}}}
  \bibinfo{year}{1973}\natexlab{}.
\newblock \showarticletitle{The theory of discrimination}.
\newblock \bibinfo{journal}{\emph{Discrimination in labor markets}}
  \bibinfo{volume}{3}, \bibinfo{number}{10} (\bibinfo{year}{1973}),
  \bibinfo{pages}{3--33}.
\newblock


\bibitem[\protect\citeauthoryear{Ashenfelter and Smith}{Ashenfelter and
  Smith}{1979}]%
        {ashenfelter1979compliance}
\bibfield{author}{\bibinfo{person}{Orley Ashenfelter} {and}
  \bibinfo{person}{Robert~S Smith}.} \bibinfo{year}{1979}\natexlab{}.
\newblock \showarticletitle{Compliance with the minimum wage law}.
\newblock \bibinfo{journal}{\emph{Journal of Political Economy}}
  \bibinfo{volume}{87}, \bibinfo{number}{2} (\bibinfo{year}{1979}),
  \bibinfo{pages}{333--350}.
\newblock


\bibitem[\protect\citeauthoryear{Barocas, Hardt, and Narayanan}{Barocas
  et~al\mbox{.}}{2019}]%
        {barocas-hardt-narayanan}
\bibfield{author}{\bibinfo{person}{Solon Barocas}, \bibinfo{person}{Moritz
  Hardt}, {and} \bibinfo{person}{Arvind Narayanan}.}
  \bibinfo{year}{2019}\natexlab{}.
\newblock \bibinfo{booktitle}{\emph{Fairness and Machine Learning}}.
\newblock \bibinfo{publisher}{fairmlbook.org}.
\newblock
\newblock
\shownote{\url{http://www.fairmlbook.org}.}


\bibitem[\protect\citeauthoryear{Barocas and Selbst}{Barocas and
  Selbst}{2016}]%
        {barocas2016big}
\bibfield{author}{\bibinfo{person}{Solon Barocas} {and}
  \bibinfo{person}{Andrew~D Selbst}.} \bibinfo{year}{2016}\natexlab{}.
\newblock \showarticletitle{Big data's disparate impact}.
\newblock \bibinfo{journal}{\emph{Calif. L. Rev.}}  \bibinfo{volume}{104}
  (\bibinfo{year}{2016}), \bibinfo{pages}{671}.
\newblock


\bibitem[\protect\citeauthoryear{{Be Applied Ltd.}}{{Be Applied Ltd.}}{2020}]%
        {applied}
\bibfield{author}{\bibinfo{person}{{Be Applied Ltd.}}}
  \bibinfo{year}{2020}\natexlab{}.
\newblock \bibinfo{title}{Why Applied?}
\newblock
\newblock
\urldef\tempurl%
\url{https://www.beapplied.com/why-applied}
\showURL{%
\tempurl}


\bibitem[\protect\citeauthoryear{Becker}{Becker}{1957}]%
        {becker1957economics}
\bibfield{author}{\bibinfo{person}{Gary~S Becker}.}
  \bibinfo{year}{1957}\natexlab{}.
\newblock \showarticletitle{The economics of discrimination Chicago}.
\newblock \bibinfo{journal}{\emph{University of Chicago}}
  (\bibinfo{year}{1957}).
\newblock


\bibitem[\protect\citeauthoryear{Bianchi and Squazzoni}{Bianchi and
  Squazzoni}{2015}]%
        {bianchi2015agent}
\bibfield{author}{\bibinfo{person}{Federico Bianchi} {and}
  \bibinfo{person}{Flaminio Squazzoni}.} \bibinfo{year}{2015}\natexlab{}.
\newblock \showarticletitle{Agent-based models in sociology}.
\newblock \bibinfo{journal}{\emph{Wiley Interdisciplinary Reviews:
  Computational Statistics}} \bibinfo{volume}{7}, \bibinfo{number}{4}
  (\bibinfo{year}{2015}), \bibinfo{pages}{284--306}.
\newblock


\bibitem[\protect\citeauthoryear{Binns}{Binns}{2018}]%
        {binns2018fairness}
\bibfield{author}{\bibinfo{person}{Reuben Binns}.}
  \bibinfo{year}{2018}\natexlab{}.
\newblock \showarticletitle{Fairness in machine learning: Lessons from
  political philosophy}. In \bibinfo{booktitle}{\emph{Conference on Fairness,
  Accountability and Transparency}}. \bibinfo{pages}{149--159}.
\newblock


\bibitem[\protect\citeauthoryear{Calders, Kamiran, and Pechenizkiy}{Calders
  et~al\mbox{.}}{2009}]%
        {Calders2009}
\bibfield{author}{\bibinfo{person}{Toon Calders}, \bibinfo{person}{Faisal
  Kamiran}, {and} \bibinfo{person}{Mykola Pechenizkiy}.}
  \bibinfo{year}{2009}\natexlab{}.
\newblock \showarticletitle{{Building Classifiers with Independency
  Constraints}}. In \bibinfo{booktitle}{\emph{Proceedings of the 2009 IEEE
  International Conference on Data Mining Workshops}}
  \emph{(\bibinfo{series}{ICDMW '09})}. \bibinfo{publisher}{IEEE Computer
  Society}, \bibinfo{address}{Washington, DC, USA}, \bibinfo{pages}{13--18}.
\newblock


\bibitem[\protect\citeauthoryear{Celis, Deshpande, Kathuria, and Vishnoi}{Celis
  et~al\mbox{.}}{2016}]%
        {celis2016fair}
\bibfield{author}{\bibinfo{person}{L~Elisa Celis}, \bibinfo{person}{Amit
  Deshpande}, \bibinfo{person}{Tarun Kathuria}, {and}
  \bibinfo{person}{Nisheeth~K Vishnoi}.} \bibinfo{year}{2016}\natexlab{}.
\newblock \showarticletitle{How to be fair and diverse?}
\newblock \bibinfo{journal}{\emph{arXiv preprint arXiv:1610.07183}}
  (\bibinfo{year}{2016}).
\newblock


\bibitem[\protect\citeauthoryear{Centola}{Centola}{2018}]%
        {centola2018behavior}
\bibfield{author}{\bibinfo{person}{Damon Centola}.}
  \bibinfo{year}{2018}\natexlab{}.
\newblock \bibinfo{booktitle}{\emph{How behavior spreads: The science of
  complex contagions}}. Vol.~\bibinfo{volume}{3}.
\newblock \bibinfo{publisher}{Princeton University Press}.
\newblock


\bibitem[\protect\citeauthoryear{Chang, Walia, et~al\mbox{.}}{Chang
  et~al\mbox{.}}{2007}]%
        {chang2007wage}
\bibfield{author}{\bibinfo{person}{Yang-Ming Chang}, \bibinfo{person}{Bhavneet
  Walia}, {et~al\mbox{.}}} \bibinfo{year}{2007}\natexlab{}.
\newblock \showarticletitle{Wage discrimination and partial compliance with the
  minimum wage law}.
\newblock \bibinfo{journal}{\emph{Economics Bulletin}} \bibinfo{volume}{10},
  \bibinfo{number}{4} (\bibinfo{year}{2007}), \bibinfo{pages}{1--7}.
\newblock


\bibitem[\protect\citeauthoryear{Chouldechova}{Chouldechova}{2017}]%
        {chouldechova2017fair}
\bibfield{author}{\bibinfo{person}{Alexandra Chouldechova}.}
  \bibinfo{year}{2017}\natexlab{}.
\newblock \showarticletitle{Fair prediction with disparate impact: A study of
  bias in recidivism prediction instruments}. In \bibinfo{booktitle}{\emph{Big
  Data}}. \bibinfo{publisher}{Mary Ann Liebert, Inc. 140 Huguenot Street, 3rd
  Floor New Rochelle, NY 10801 USA}.
\newblock


\bibitem[\protect\citeauthoryear{Coate and Loury}{Coate and Loury}{1993}]%
        {coate1993will}
\bibfield{author}{\bibinfo{person}{Stephen Coate} {and}
  \bibinfo{person}{Glenn~C Loury}.} \bibinfo{year}{1993}\natexlab{}.
\newblock \showarticletitle{Will affirmative-action policies eliminate negative
  stereotypes?}
\newblock \bibinfo{journal}{\emph{The American Economic Review}}
  (\bibinfo{year}{1993}), \bibinfo{pages}{1220--1240}.
\newblock


\bibitem[\protect\citeauthoryear{D'Amour, Srinivasan, Atwood, Baljekar,
  Sculley, and Halpern}{D'Amour et~al\mbox{.}}{2020}]%
        {d2020fairness}
\bibfield{author}{\bibinfo{person}{Alexander D'Amour}, \bibinfo{person}{Hansa
  Srinivasan}, \bibinfo{person}{James Atwood}, \bibinfo{person}{Pallavi
  Baljekar}, \bibinfo{person}{D Sculley}, {and} \bibinfo{person}{Yoni
  Halpern}.} \bibinfo{year}{2020}\natexlab{}.
\newblock \showarticletitle{Fairness is not static: deeper understanding of
  long term fairness via simulation studies}. In
  \bibinfo{booktitle}{\emph{Proceedings of the 2020 Conference on Fairness,
  Accountability, and Transparency}}. \bibinfo{pages}{525--534}.
\newblock


\bibitem[\protect\citeauthoryear{Dieterich, Mendoza, and Brennan}{Dieterich
  et~al\mbox{.}}{2016}]%
        {dieterich2016compas}
\bibfield{author}{\bibinfo{person}{William Dieterich},
  \bibinfo{person}{Christina Mendoza}, {and} \bibinfo{person}{Tim Brennan}.}
  \bibinfo{year}{2016}\natexlab{}.
\newblock \showarticletitle{COMPAS risk scales: Demonstrating accuracy equity
  and predictive parity}.
\newblock \bibinfo{journal}{\emph{Northpointe Inc}} (\bibinfo{year}{2016}).
\newblock


\bibitem[\protect\citeauthoryear{Drosou, Jagadish, Pitoura, and
  Stoyanovich}{Drosou et~al\mbox{.}}{2017}]%
        {drosou2017diversity}
\bibfield{author}{\bibinfo{person}{Marina Drosou}, \bibinfo{person}{HV
  Jagadish}, \bibinfo{person}{Evaggelia Pitoura}, {and} \bibinfo{person}{Julia
  Stoyanovich}.} \bibinfo{year}{2017}\natexlab{}.
\newblock \showarticletitle{Diversity in big data: A review}.
\newblock \bibinfo{journal}{\emph{Big data}} \bibinfo{volume}{5},
  \bibinfo{number}{2} (\bibinfo{year}{2017}), \bibinfo{pages}{73--84}.
\newblock


\bibitem[\protect\citeauthoryear{Elster}{Elster}{1992}]%
        {elster1992local}
\bibfield{author}{\bibinfo{person}{Jon Elster}.}
  \bibinfo{year}{1992}\natexlab{}.
\newblock \bibinfo{booktitle}{\emph{Local justice: How institutions allocate
  scarce goods and necessary burdens}}.
\newblock \bibinfo{publisher}{Russell Sage Foundation}.
\newblock


\bibitem[\protect\citeauthoryear{Fazelpour and Lipton}{Fazelpour and
  Lipton}{2020}]%
        {fazelpour2020algorithmic}
\bibfield{author}{\bibinfo{person}{Sina Fazelpour} {and}
  \bibinfo{person}{Zachary~C Lipton}.} \bibinfo{year}{2020}\natexlab{}.
\newblock \showarticletitle{Algorithmic Fairness from a Non-ideal Perspective}.
  In \bibinfo{booktitle}{\emph{Proceedings of the AAAI/ACM Conference on AI,
  Ethics, and Society}}. \bibinfo{pages}{57--63}.
\newblock


\bibitem[\protect\citeauthoryear{Feldman, Friedler, Moeller, Scheidegger, and
  Venkatasubramanian}{Feldman et~al\mbox{.}}{2015}]%
        {feldman2015certifying}
\bibfield{author}{\bibinfo{person}{Michael Feldman}, \bibinfo{person}{Sorelle~A
  Friedler}, \bibinfo{person}{John Moeller}, \bibinfo{person}{Carlos
  Scheidegger}, {and} \bibinfo{person}{Suresh Venkatasubramanian}.}
  \bibinfo{year}{2015}\natexlab{}.
\newblock \showarticletitle{Certifying and removing disparate impact}. In
  \bibinfo{booktitle}{\emph{proceedings of the 21th ACM SIGKDD international
  conference on knowledge discovery and data mining}}.
  \bibinfo{pages}{259--268}.
\newblock


\bibitem[\protect\citeauthoryear{Fullinwider}{Fullinwider}{2018}]%
        {sep-affirmative-action}
\bibfield{author}{\bibinfo{person}{Robert Fullinwider}.}
  \bibinfo{year}{2018}\natexlab{}.
\newblock \showarticletitle{{Affirmative Action}}.
\newblock In \bibinfo{booktitle}{\emph{The {Stanford} Encyclopedia of
  Philosophy} (\bibinfo{edition}{summer 2018} ed.)},
  \bibfield{editor}{\bibinfo{person}{Edward~N. Zalta}} (Ed.).
  \bibinfo{publisher}{Metaphysics Research Lab, Stanford University}.
\newblock


\bibitem[\protect\citeauthoryear{Green and Chen}{Green and Chen}{2019}]%
        {green2019disparate}
\bibfield{author}{\bibinfo{person}{Ben Green} {and} \bibinfo{person}{Yiling
  Chen}.} \bibinfo{year}{2019}\natexlab{}.
\newblock \showarticletitle{Disparate interactions: An algorithm-in-the-loop
  analysis of fairness in risk assessments}. In
  \bibinfo{booktitle}{\emph{Proceedings of the Conference on Fairness,
  Accountability, and Transparency}}. \bibinfo{pages}{90--99}.
\newblock


\bibitem[\protect\citeauthoryear{Green and Hu}{Green and Hu}{2018}]%
        {green2018myth}
\bibfield{author}{\bibinfo{person}{Ben Green} {and} \bibinfo{person}{Lily Hu}.}
  \bibinfo{year}{2018}\natexlab{}.
\newblock \showarticletitle{The myth in the methodology: Towards a
  recontextualization of fairness in machine learning}. In
  \bibinfo{booktitle}{\emph{Proceedings of the machine learning: the debates
  workshop}}.
\newblock


\bibitem[\protect\citeauthoryear{Grunwald}{Grunwald}{2014}]%
        {grunwald2014technology}
\bibfield{author}{\bibinfo{person}{Armin Grunwald}.}
  \bibinfo{year}{2014}\natexlab{}.
\newblock \showarticletitle{Technology assessment for responsible innovation}.
\newblock In \bibinfo{booktitle}{\emph{Responsible Innovation 1}}.
  \bibinfo{publisher}{Springer}, \bibinfo{pages}{15--31}.
\newblock


\bibitem[\protect\citeauthoryear{Hansson}{Hansson}{2013}]%
        {hansson2013ethics}
\bibfield{author}{\bibinfo{person}{S Hansson}.}
  \bibinfo{year}{2013}\natexlab{}.
\newblock \bibinfo{booktitle}{\emph{The ethics of risk: Ethical analysis in an
  uncertain world}}.
\newblock \bibinfo{publisher}{Springer}.
\newblock


\bibitem[\protect\citeauthoryear{Hardt, Megiddo, Papadimitriou, and
  Wootters}{Hardt et~al\mbox{.}}{2016}]%
        {hardt2016strategic}
\bibfield{author}{\bibinfo{person}{Moritz Hardt}, \bibinfo{person}{Nimrod
  Megiddo}, \bibinfo{person}{Christos Papadimitriou}, {and}
  \bibinfo{person}{Mary Wootters}.} \bibinfo{year}{2016}\natexlab{}.
\newblock \showarticletitle{Strategic classification}. In
  \bibinfo{booktitle}{\emph{Proceedings of the 2016 ACM conference on
  innovations in theoretical computer science}}. \bibinfo{pages}{111--122}.
\newblock


\bibitem[\protect\citeauthoryear{Heidari, Nanda, and Gummadi}{Heidari
  et~al\mbox{.}}{2019}]%
        {heidari2019long}
\bibfield{author}{\bibinfo{person}{Hoda Heidari}, \bibinfo{person}{Vedant
  Nanda}, {and} \bibinfo{person}{Krishna~P Gummadi}.}
  \bibinfo{year}{2019}\natexlab{}.
\newblock \showarticletitle{On the long-term impact of algorithmic decision
  policies: Effort unfairness and feature segregation through social learning}.
\newblock \bibinfo{journal}{\emph{arXiv preprint arXiv:1903.01209}}
  (\bibinfo{year}{2019}).
\newblock


\bibitem[\protect\citeauthoryear{Hu and Chen}{Hu and Chen}{2018}]%
        {hu2018short}
\bibfield{author}{\bibinfo{person}{Lily Hu} {and} \bibinfo{person}{Yiling
  Chen}.} \bibinfo{year}{2018}\natexlab{}.
\newblock \showarticletitle{A short-term intervention for long-term fairness in
  the labor market}. In \bibinfo{booktitle}{\emph{Proceedings of the 2018 World
  Wide Web Conference}}. \bibinfo{pages}{1389--1398}.
\newblock


\bibitem[\protect\citeauthoryear{Hu, Immorlica, and Vaughan}{Hu
  et~al\mbox{.}}{2018}]%
        {hu2018disparate}
\bibfield{author}{\bibinfo{person}{Lily Hu}, \bibinfo{person}{Nicole
  Immorlica}, {and} \bibinfo{person}{Jennifer~Wortman Vaughan}.}
  \bibinfo{year}{2018}\natexlab{}.
\newblock \showarticletitle{The Disparate Effects of Strategic Manipulation}.
  In \bibinfo{booktitle}{\emph{Conference on Fairness Accountability and
  Transparency (FAT*)}}.
\newblock


\bibitem[\protect\citeauthoryear{Hu, Wu, Zhang, and Wu}{Hu
  et~al\mbox{.}}{2020}]%
        {hu2020fair}
\bibfield{author}{\bibinfo{person}{Yaowei Hu}, \bibinfo{person}{Yongkai Wu},
  \bibinfo{person}{Lu Zhang}, {and} \bibinfo{person}{Xintao Wu}.}
  \bibinfo{year}{2020}\natexlab{}.
\newblock \showarticletitle{Fair Multiple Decision Making Through Soft
  Interventions}.
\newblock \bibinfo{journal}{\emph{Advances in Neural Information Processing
  Systems}}  \bibinfo{volume}{33} (\bibinfo{year}{2020}).
\newblock


\bibitem[\protect\citeauthoryear{Kasy and Abebe}{Kasy and Abebe}{2020}]%
        {kasy2020fairness}
\bibfield{author}{\bibinfo{person}{Maximilian Kasy} {and}
  \bibinfo{person}{Rediet Abebe}.} \bibinfo{year}{2020}\natexlab{}.
\newblock \bibinfo{booktitle}{\emph{Fairness, equality, and power in
  algorithmic decision making}}.
\newblock \bibinfo{type}{{T}echnical {R}eport}. \bibinfo{institution}{Working
  paper}.
\newblock


\bibitem[\protect\citeauthoryear{Kiesling, G{\"u}nther, Stummer, and
  Wakolbinger}{Kiesling et~al\mbox{.}}{2012}]%
        {kiesling2012agent}
\bibfield{author}{\bibinfo{person}{Elmar Kiesling}, \bibinfo{person}{Markus
  G{\"u}nther}, \bibinfo{person}{Christian Stummer}, {and}
  \bibinfo{person}{Lea~M Wakolbinger}.} \bibinfo{year}{2012}\natexlab{}.
\newblock \showarticletitle{Agent-based simulation of innovation diffusion: a
  review}.
\newblock \bibinfo{journal}{\emph{Central European Journal of Operations
  Research}} \bibinfo{volume}{20}, \bibinfo{number}{2} (\bibinfo{year}{2012}),
  \bibinfo{pages}{183--230}.
\newblock


\bibitem[\protect\citeauthoryear{Kilbertus, Rodriguez, Sch{\"o}lkopf, Muandet,
  and Valera}{Kilbertus et~al\mbox{.}}{2020}]%
        {kilbertus2020fair}
\bibfield{author}{\bibinfo{person}{Niki Kilbertus},
  \bibinfo{person}{Manuel~Gomez Rodriguez}, \bibinfo{person}{Bernhard
  Sch{\"o}lkopf}, \bibinfo{person}{Krikamol Muandet}, {and}
  \bibinfo{person}{Isabel Valera}.} \bibinfo{year}{2020}\natexlab{}.
\newblock \showarticletitle{Fair decisions despite imperfect predictions}. In
  \bibinfo{booktitle}{\emph{International Conference on Artificial Intelligence
  and Statistics}}. PMLR, \bibinfo{pages}{277--287}.
\newblock


\bibitem[\protect\citeauthoryear{Kleinberg and Raghavan}{Kleinberg and
  Raghavan}{2019}]%
        {kleinberg2019classifiers}
\bibfield{author}{\bibinfo{person}{Jon Kleinberg} {and} \bibinfo{person}{Manish
  Raghavan}.} \bibinfo{year}{2019}\natexlab{}.
\newblock \showarticletitle{How Do Classifiers Induce Agents to Invest Effort
  Strategically?}. In \bibinfo{booktitle}{\emph{ACM Conference on Economics and
  Computation (EC)}}.
\newblock


\bibitem[\protect\citeauthoryear{Lipton, McAuley, and Chouldechova}{Lipton
  et~al\mbox{.}}{2018}]%
        {lipton2018does}
\bibfield{author}{\bibinfo{person}{Zachary Lipton}, \bibinfo{person}{Julian
  McAuley}, {and} \bibinfo{person}{Alexandra Chouldechova}.}
  \bibinfo{year}{2018}\natexlab{}.
\newblock \showarticletitle{Does mitigating ML's impact disparity require
  treatment disparity?}. In \bibinfo{booktitle}{\emph{Advances in Neural
  Information Processing Systems}}. \bibinfo{pages}{8125--8135}.
\newblock


\bibitem[\protect\citeauthoryear{Lipton and Steinhardt}{Lipton and
  Steinhardt}{2018}]%
        {Lipton2018a}
\bibfield{author}{\bibinfo{person}{Zachary~C. Lipton} {and}
  \bibinfo{person}{Jacob Steinhardt}.} \bibinfo{year}{2018}\natexlab{}.
\newblock \showarticletitle{{Troubling Trends in Machine Learning
  Scholarship}}.
\newblock \bibinfo{journal}{\emph{Communications of the ACM (CACM)}}
  \bibinfo{volume}{62}, \bibinfo{number}{6} (\bibinfo{year}{2018}),
  \bibinfo{pages}{45--53}.
\newblock


\bibitem[\protect\citeauthoryear{Liu, Dean, Rolf, Simchowitz, and Hardt}{Liu
  et~al\mbox{.}}{2019}]%
        {liu2019delayed}
\bibfield{author}{\bibinfo{person}{Lydia~T Liu}, \bibinfo{person}{Sarah Dean},
  \bibinfo{person}{Esther Rolf}, \bibinfo{person}{Max Simchowitz}, {and}
  \bibinfo{person}{Moritz Hardt}.} \bibinfo{year}{2019}\natexlab{}.
\newblock \showarticletitle{Delayed impact of fair machine learning}. In
  \bibinfo{booktitle}{\emph{Proceedings of the 28th International Joint
  Conference on Artificial Intelligence}}. AAAI Press,
  \bibinfo{pages}{6196--6200}.
\newblock


\bibitem[\protect\citeauthoryear{Liu, Wilson, Haghtalab, Kalai, Borgs, and
  Chayes}{Liu et~al\mbox{.}}{2020}]%
        {liu2020disparate}
\bibfield{author}{\bibinfo{person}{Lydia~T Liu}, \bibinfo{person}{Ashia
  Wilson}, \bibinfo{person}{Nika Haghtalab}, \bibinfo{person}{Adam~Tauman
  Kalai}, \bibinfo{person}{Christian Borgs}, {and} \bibinfo{person}{Jennifer
  Chayes}.} \bibinfo{year}{2020}\natexlab{}.
\newblock \showarticletitle{The disparate equilibria of algorithmic decision
  making when individuals invest rationally}. In
  \bibinfo{booktitle}{\emph{Proceedings of the 2020 Conference on Fairness,
  Accountability, and Transparency}}. \bibinfo{pages}{381--391}.
\newblock


\bibitem[\protect\citeauthoryear{Miller}{Miller}{2011}]%
        {Miller2011-MILTUT-3}
\bibfield{author}{\bibinfo{person}{David Miller}.}
  \bibinfo{year}{2011}\natexlab{}.
\newblock \showarticletitle{Taking Up the Slack? Responsibility and Justice in
  Situations of Partial Compliance}.
\newblock In \bibinfo{booktitle}{\emph{Responsibility and Distributive
  Justice}}, \bibfield{editor}{\bibinfo{person}{Carl Knight} {and}
  \bibinfo{person}{Zofia Stemplowska}} (Eds.). \bibinfo{publisher}{Oxford
  University Press}, \bibinfo{pages}{230--45}.
\newblock


\bibitem[\protect\citeauthoryear{Milli, Miller, Dragan, and Hardt}{Milli
  et~al\mbox{.}}{2018}]%
        {milli2018social}
\bibfield{author}{\bibinfo{person}{Smitha Milli}, \bibinfo{person}{John
  Miller}, \bibinfo{person}{Anca~D Dragan}, {and} \bibinfo{person}{Moritz
  Hardt}.} \bibinfo{year}{2018}\natexlab{}.
\newblock \showarticletitle{The Social Cost of Strategic Classification}. In
  \bibinfo{booktitle}{\emph{Conference on Fairness Accountability and
  Transparency (FAT*)}}.
\newblock


\bibitem[\protect\citeauthoryear{Muldoon}{Muldoon}{2016}]%
        {muldoon2016social}
\bibfield{author}{\bibinfo{person}{Ryan Muldoon}.}
  \bibinfo{year}{2016}\natexlab{}.
\newblock \bibinfo{booktitle}{\emph{Social contract theory for a diverse world:
  Beyond tolerance}}.
\newblock \bibinfo{publisher}{Taylor \& Francis}.
\newblock


\bibitem[\protect\citeauthoryear{O'Connor}{O'Connor}{2019}]%
        {OConnor2019a}
\bibfield{author}{\bibinfo{person}{Cailin O'Connor}.}
  \bibinfo{year}{2019}\natexlab{}.
\newblock \bibinfo{booktitle}{\emph{{The Origins of Unfairness}}}.
\newblock \bibinfo{publisher}{Oxford University Press},
  \bibinfo{address}{Oxford}.
\newblock


\bibitem[\protect\citeauthoryear{Page}{Page}{2019}]%
        {page2019diversity}
\bibfield{author}{\bibinfo{person}{Scott~E Page}.}
  \bibinfo{year}{2019}\natexlab{}.
\newblock \bibinfo{booktitle}{\emph{The diversity bonus: How great teams pay
  off in the knowledge economy}}.
\newblock \bibinfo{publisher}{Princeton University Press}.
\newblock


\bibitem[\protect\citeauthoryear{Parikh, Obermeyer, and Navathe}{Parikh
  et~al\mbox{.}}{2019}]%
        {parikh2019regulation}
\bibfield{author}{\bibinfo{person}{Ravi~B Parikh}, \bibinfo{person}{Ziad
  Obermeyer}, {and} \bibinfo{person}{Amol~S Navathe}.}
  \bibinfo{year}{2019}\natexlab{}.
\newblock \showarticletitle{Regulation of predictive analytics in medicine}.
\newblock \bibinfo{journal}{\emph{Science}} \bibinfo{volume}{363},
  \bibinfo{number}{6429} (\bibinfo{year}{2019}), \bibinfo{pages}{810--812}.
\newblock


\bibitem[\protect\citeauthoryear{Phelps}{Phelps}{1972}]%
        {phelps1972statistical}
\bibfield{author}{\bibinfo{person}{Edmund~S Phelps}.}
  \bibinfo{year}{1972}\natexlab{}.
\newblock \showarticletitle{The statistical theory of racism and sexism}.
\newblock \bibinfo{journal}{\emph{The american economic review}}
  \bibinfo{volume}{62}, \bibinfo{number}{4} (\bibinfo{year}{1972}),
  \bibinfo{pages}{659--661}.
\newblock


\bibitem[\protect\citeauthoryear{Pleiss, Raghavan, Wu, Kleinberg, and
  Weinberger}{Pleiss et~al\mbox{.}}{2017}]%
        {pleiss2017fairness}
\bibfield{author}{\bibinfo{person}{Geoff Pleiss}, \bibinfo{person}{Manish
  Raghavan}, \bibinfo{person}{Felix Wu}, \bibinfo{person}{Jon Kleinberg}, {and}
  \bibinfo{person}{Kilian~Q Weinberger}.} \bibinfo{year}{2017}\natexlab{}.
\newblock \showarticletitle{On Fairness and Calibration}.
\newblock \bibinfo{journal}{\emph{Advances in Neural Information Processing
  Systems}}  \bibinfo{volume}{30} (\bibinfo{year}{2017}),
  \bibinfo{pages}{5680--5689}.
\newblock


\bibitem[\protect\citeauthoryear{Raji, Smart, White, Mitchell, Gebru,
  Hutchinson, Smith-Loud, Theron, and Barnes}{Raji et~al\mbox{.}}{2020}]%
        {raji2020closing}
\bibfield{author}{\bibinfo{person}{Inioluwa~Deborah Raji},
  \bibinfo{person}{Andrew Smart}, \bibinfo{person}{Rebecca~N White},
  \bibinfo{person}{Margaret Mitchell}, \bibinfo{person}{Timnit Gebru},
  \bibinfo{person}{Ben Hutchinson}, \bibinfo{person}{Jamila Smith-Loud},
  \bibinfo{person}{Daniel Theron}, {and} \bibinfo{person}{Parker Barnes}.}
  \bibinfo{year}{2020}\natexlab{}.
\newblock \showarticletitle{Closing the AI accountability gap: defining an
  end-to-end framework for internal algorithmic auditing}. In
  \bibinfo{booktitle}{\emph{Proceedings of the 2020 Conference on Fairness,
  Accountability, and Transparency}}. \bibinfo{pages}{33--44}.
\newblock


\bibitem[\protect\citeauthoryear{Rambachan, Kleinberg, Mullainathan, and
  Ludwig}{Rambachan et~al\mbox{.}}{2020}]%
        {rambachan2020economic}
\bibfield{author}{\bibinfo{person}{Ashesh Rambachan}, \bibinfo{person}{Jon
  Kleinberg}, \bibinfo{person}{Sendhil Mullainathan}, {and}
  \bibinfo{person}{Jens Ludwig}.} \bibinfo{year}{2020}\natexlab{}.
\newblock \bibinfo{booktitle}{\emph{An economic approach to regulating
  algorithms}}.
\newblock \bibinfo{type}{{T}echnical {R}eport}. \bibinfo{institution}{National
  Bureau of Economic Research}.
\newblock


\bibitem[\protect\citeauthoryear{S{\'a}nchez-Monedero, Dencik, and
  Edwards}{S{\'a}nchez-Monedero et~al\mbox{.}}{2020}]%
        {sanchez2020does}
\bibfield{author}{\bibinfo{person}{Javier S{\'a}nchez-Monedero},
  \bibinfo{person}{Lina Dencik}, {and} \bibinfo{person}{Lilian Edwards}.}
  \bibinfo{year}{2020}\natexlab{}.
\newblock \showarticletitle{What does it mean to'solve'the problem of
  discrimination in hiring? social, technical and legal perspectives from the
  UK on automated hiring systems}. In \bibinfo{booktitle}{\emph{Proceedings of
  the 2020 Conference on Fairness, Accountability, and Transparency}}.
  \bibinfo{pages}{458--468}.
\newblock


\bibitem[\protect\citeauthoryear{Sandvig, Hamilton, Karahalios, and
  Langbort}{Sandvig et~al\mbox{.}}{2014}]%
        {sandvig2014auditing}
\bibfield{author}{\bibinfo{person}{Christian Sandvig}, \bibinfo{person}{Kevin
  Hamilton}, \bibinfo{person}{Karrie Karahalios}, {and} \bibinfo{person}{Cedric
  Langbort}.} \bibinfo{year}{2014}\natexlab{}.
\newblock \showarticletitle{Auditing algorithms: Research methods for detecting
  discrimination on internet platforms}.
\newblock \bibinfo{journal}{\emph{Data and discrimination: converting critical
  concerns into productive inquiry}}  \bibinfo{volume}{22}
  (\bibinfo{year}{2014}), \bibinfo{pages}{4349--4357}.
\newblock


\bibitem[\protect\citeauthoryear{Schapiro}{Schapiro}{2003}]%
        {schapiro2003compliance}
\bibfield{author}{\bibinfo{person}{Tamar Schapiro}.}
  \bibinfo{year}{2003}\natexlab{}.
\newblock \showarticletitle{Compliance, complicity, and the nature of nonideal
  conditions}.
\newblock \bibinfo{journal}{\emph{The Journal of Philosophy}}
  (\bibinfo{year}{2003}).
\newblock


\bibitem[\protect\citeauthoryear{Selbst, Boyd, Friedler, Venkatasubramanian,
  and Vertesi}{Selbst et~al\mbox{.}}{2019}]%
        {Selbst2019}
\bibfield{author}{\bibinfo{person}{Andrew~D. Selbst}, \bibinfo{person}{Danah
  Boyd}, \bibinfo{person}{Sorelle~A. Friedler}, \bibinfo{person}{Suresh
  Venkatasubramanian}, {and} \bibinfo{person}{Janet Vertesi}.}
  \bibinfo{year}{2019}\natexlab{}.
\newblock \showarticletitle{Fairness and Abstraction in Sociotechnical
  Systems}. In \bibinfo{booktitle}{\emph{Fairness, Accountability, and
  Transparency (FAT*)}}.
\newblock


\bibitem[\protect\citeauthoryear{Smith-Doerr, Alegria, and Sacco}{Smith-Doerr
  et~al\mbox{.}}{2017}]%
        {smith2017diversity}
\bibfield{author}{\bibinfo{person}{Laurel Smith-Doerr},
  \bibinfo{person}{Sharla~N Alegria}, {and} \bibinfo{person}{Timothy Sacco}.}
  \bibinfo{year}{2017}\natexlab{}.
\newblock \showarticletitle{How diversity matters in the US science and
  engineering workforce: A critical review considering integration in teams,
  fields, and organizational contexts}.
\newblock \bibinfo{journal}{\emph{Engaging Science, Technology, and Society}}
  \bibinfo{volume}{3} (\bibinfo{year}{2017}), \bibinfo{pages}{139--153}.
\newblock


\bibitem[\protect\citeauthoryear{Squire and Suthiwart-Narueput}{Squire and
  Suthiwart-Narueput}{1997}]%
        {squire1997impact}
\bibfield{author}{\bibinfo{person}{Lyn Squire} {and} \bibinfo{person}{Sethaput
  Suthiwart-Narueput}.} \bibinfo{year}{1997}\natexlab{}.
\newblock \showarticletitle{The Impact of Labor Market Regulations}.
\newblock \bibinfo{journal}{\emph{The World Bank Economic Review}}
  (\bibinfo{year}{1997}), \bibinfo{pages}{119--143}.
\newblock


\bibitem[\protect\citeauthoryear{Steel, Fazelpour, Crewe, and Gillette}{Steel
  et~al\mbox{.}}{2019}]%
        {steel2019information}
\bibfield{author}{\bibinfo{person}{Daniel Steel}, \bibinfo{person}{Sina
  Fazelpour}, \bibinfo{person}{Bianca Crewe}, {and} \bibinfo{person}{Kinley
  Gillette}.} \bibinfo{year}{2019}\natexlab{}.
\newblock \showarticletitle{Information elaboration and epistemic effects of
  diversity}.
\newblock \bibinfo{journal}{\emph{Synthese}} (\bibinfo{year}{2019}),
  \bibinfo{pages}{1--21}.
\newblock


\bibitem[\protect\citeauthoryear{Steel, Fazelpour, Gillette, Crewe, and
  Burgess}{Steel et~al\mbox{.}}{2018}]%
        {steel2018multiple}
\bibfield{author}{\bibinfo{person}{Daniel Steel}, \bibinfo{person}{Sina
  Fazelpour}, \bibinfo{person}{Kinley Gillette}, \bibinfo{person}{Bianca
  Crewe}, {and} \bibinfo{person}{Michael Burgess}.}
  \bibinfo{year}{2018}\natexlab{}.
\newblock \showarticletitle{Multiple diversity concepts and their
  ethical-epistemic implications}.
\newblock \bibinfo{journal}{\emph{European Journal for Philosophy of Science}}
  \bibinfo{volume}{8}, \bibinfo{number}{3} (\bibinfo{year}{2018}),
  \bibinfo{pages}{761--780}.
\newblock


\bibitem[\protect\citeauthoryear{Thomson}{Thomson}{1973}]%
        {thomson1973preferential}
\bibfield{author}{\bibinfo{person}{Judith~Jarvis Thomson}.}
  \bibinfo{year}{1973}\natexlab{}.
\newblock \showarticletitle{Preferential hiring}.
\newblock \bibinfo{journal}{\emph{Philosophy \& Public Affairs}}
  (\bibinfo{year}{1973}), \bibinfo{pages}{364--384}.
\newblock


\bibitem[\protect\citeauthoryear{Valentini}{Valentini}{2012}]%
        {Valentini2012}
\bibfield{author}{\bibinfo{person}{Laura Valentini}.}
  \bibinfo{year}{2012}\natexlab{}.
\newblock \showarticletitle{{Ideal vs. Non-ideal Theory: A Conceptual Map}}.
\newblock \bibinfo{journal}{\emph{Philosophy Compass}} (\bibinfo{year}{2012}).
\newblock


\bibitem[\protect\citeauthoryear{Wachter, Mittelstadt, and Russell}{Wachter
  et~al\mbox{.}}{2020}]%
        {wachter2020fairness}
\bibfield{author}{\bibinfo{person}{Sandra Wachter}, \bibinfo{person}{Brent
  Mittelstadt}, {and} \bibinfo{person}{Chris Russell}.}
  \bibinfo{year}{2020}\natexlab{}.
\newblock \showarticletitle{Why fairness cannot be automated: Bridging the gap
  between EU non-discrimination law and AI}.
\newblock \bibinfo{journal}{\emph{Available at SSRN}} (\bibinfo{year}{2020}).
\newblock


\bibitem[\protect\citeauthoryear{Warren}{Warren}{1977}]%
        {warren1977secondary}
\bibfield{author}{\bibinfo{person}{Mary~Anne Warren}.}
  \bibinfo{year}{1977}\natexlab{}.
\newblock \showarticletitle{Secondary sexism and quota hiring}.
\newblock \bibinfo{journal}{\emph{Philosophy \& Public Affairs}}
  (\bibinfo{year}{1977}), \bibinfo{pages}{240--261}.
\newblock


\bibitem[\protect\citeauthoryear{Wilson, Ghosh, Jiang, Mislove, Baker, Szary,
  Trindel, and Polli}{Wilson et~al\mbox{.}}{[n.d.]}]%
        {wilsonbuilding}
\bibfield{author}{\bibinfo{person}{Christo Wilson}, \bibinfo{person}{Avijit
  Ghosh}, \bibinfo{person}{Shan Jiang}, \bibinfo{person}{Alan Mislove},
  \bibinfo{person}{Lewis Baker}, \bibinfo{person}{Janelle Szary},
  \bibinfo{person}{Kelly Trindel}, {and} \bibinfo{person}{Frida Polli}.}
  \bibinfo{year}{[n.d.]}\natexlab{}.
\newblock \showarticletitle{Building and Auditing Fair Algorithms: A Case Study
  in Candidate Screening}.
\newblock  (\bibinfo{year}{[n.\,d.]}).
\newblock


\bibitem[\protect\citeauthoryear{Zollman}{Zollman}{2013}]%
        {zollman2013network}
\bibfield{author}{\bibinfo{person}{Kevin~JS Zollman}.}
  \bibinfo{year}{2013}\natexlab{}.
\newblock \showarticletitle{Network epistemology: Communication in epistemic
  communities}.
\newblock \bibinfo{journal}{\emph{Philosophy Compass}} \bibinfo{volume}{8},
  \bibinfo{number}{1} (\bibinfo{year}{2013}), \bibinfo{pages}{15--27}.
\newblock


\end{thebibliography}

\end{document}